\documentclass{Interspeech}
\usepackage{tcolorbox}
\usepackage{graphicx} 

\interspeechcameraready

\title{MultiActor-Audiobook: Zero-Shot Audiobook Generation \\with Faces and Voices of  Multiple Speakers}

\author[affiliation={1}]{Kyeongman}{Park}
\author[affiliation={1}]{Seongho}{Joo}
\author[affiliation={1}]{Kyomin}{Jung}

\affiliation{Seoul National University}{South Korea}{}
\email{zzangmane@snu.ac.kr, seonghojoo@snu.ac.kr, kjung@snu.ac.kr}


\usepackage{comment}

\begin{document}

\maketitle

\begin{abstract}
    We introduce MultiActor-Audiobook, a zero-shot approach for generating audiobooks that automatically produces consistent, expressive, and speaker-appropriate prosody, including intonation and emotion. Previous audiobook systems have several limitations: they require users to manually configure the speaker's prosody, read each sentence with a monotonic tone compared to voice actors, or rely on costly training. However, our MultiActor-Audiobook addresses these issues by introducing two novel processes: (1) MSP (\textbf{Multimodal Speaker Persona Generation}) and (2) LSI (\textbf{LLM-based Script Instruction Generation}). With these two processes, MultiActor-Audiobook can generate more emotionally expressive audiobooks with a consistent speaker prosody without additional training. We compare our system with commercial products, through human and MLLM evaluations, achieving competitive results. Furthermore, we demonstrate the effectiveness of MSP and LSI through ablation studies.
\end{abstract}

\section{Introduction}
Stories are complex literary works characterized by a wide range of emotions and multiple speakers. Transforming such texts into speech, as in the case of audiobooks, resembles human actors performing a script, where they convey the most fitting prosody and emotions for each scene \cite{ge2019text,vyas2023audiobox,ramli2016rule}. For instance, reading a novel in a monotone voice would be unnatural; instead, the emotion, tone, and pitch must dynamically adapt to the context \cite{wei2023discourse,guo2023prompttts,leng2023prompttts,shimizu2024prompttts++,diatlova2023emospeech}. Furthermore, stories feature not only narrators but also diverse characters, each requiring a voice that reflects their distinct characteristics such as personality and physical traits like face \cite{peng2024voicecraft,le2024voicebox,eskimez2024e2,guan2024mm,lee2024hear,li2025flespeech,zhang2022transformer,fu2025vita} to achieve natural-sounding dialogue.

Traditional audiobook generation systems have generally relied on two approaches. The first approach involves collecting extensive, high-cost datasets, such as over 60 hours of professional voice actor recordings \cite{ge2019text}, to predict dynamic prosody for each word embedding\cite{ge2019text,bae2023sound,pethe2023prosody}. However, this method is limited to the specific domain in which the data was collected and incurs significant costs for data acquisition and model training. For example, \cite{ge2019text} can only produce natural speech within the scope of traditional Chinese oral art forms.

The second approach is manual annotating, where humans meticulously select cadence, emotions, pitches, and tones for each speaker and each sentences \cite{ramli2016rule,zhao2023narrativeplay,zhang2023expressive}. This annotated data is then used with pre-trained prompt-TTS systems \cite{vyas2023audiobox,guo2023prompttts,leng2023prompttts,shimizu2024prompttts++,guan2024mm,lee2024fvtts}. While this method reduces costs for data collection and model training, it still requires substantial human labor and time to produce a single audiobook. For instance, NarrativePlay \cite{zhao2023narrativeplay} involved manually selecting from over 3,000 predefined prosody patterns for each character.

To sum up, previous works often need costly data collection, model training, or manual annotations. To address these challenges, we propose \textbf{MultiActor-Audiobook} : \textit{Zero-Shot Audiobook Generation with Faces and Voices of  Multiple Speakers}. MultiActor-Audiobook introduces two innovative processes: (1) \textbf{Multimodal Speaker Persona Generation} and (2) \textbf{LLM-Based Script Instruction Generation}. Although both processes operate in a fully automated and zero-shot manner, MultiActor-Audiobook produces more emotionally expressive and speaker-appropriate audiobooks.

During Multimodal Speaker Persona Generation, we create a multimodal persona for each speaker. Specifically, LLM first identifies all speaker entities within the novel, including both dialogue and narration. It then extracts descriptive features from the text related to each character. Based on this textual persona information, we utilize a text2image model to generate an AI-generated face image that visually represents the character. Using this face image and its corresponding caption, we employ a pretrained Face-to-Voice system \cite{li2025flespeech} to produce a unique voice sample that reflects each character's distinctive prosody. By leveraging visual information for each character, we can generate a more characteristic and fitting voice, and we can maintain speaker consistency throughout the story by anchoring each speaker to their unique voice samples.

Then the LLM-based Script Instruction Generation process employs GPT-4o \cite{achiam2023gpt} to create dynamic instructions for each sentence in the novel. These instructions mainly include emotional cues, tone, and pitch. During generation, we not only provide the target sentence but also give the surrounding context and each character's persona information to LLM. By providing this additional information, the LLM can predict more appropriate and continual emotional expression for each sentence. With these detailed emotional and contextual script instructions, the TTS model can read each sentence more naturally and expressively, leading to a better listening experience.

To evaluate our system's performance, human annotators and the MLLM (Multi-modal Large Language Model) compare it with baselines, including powerful commercial products. As a result, our system achieves comparable MOS scores to the baselines in human evaluation and demonstrates an average 0.225-point improvement in MLLM evaluation. Additionally, our ablation studies validate the effectiveness of each process by showing consistent improvements across all metrics, highlighting its respective contribution to overall performance.


\section{MultiActor-Audiobook}
MultiActor-Audiobook aims to create audiobooks that deliver speaker-aligned voice and natural emotional expressions. To generate each speaker-algined voices in the story, it performs \textbf{Multimodal Speaker Persona Generation}, and to ensure natural emotional expression, it utilizes \textbf{LLM-Based Script Instruction Generation}.
\subsection{Multimodal Speaker Persona Generation}
In this process, the LLM identifies each character in the story and generates audio samples and facial images that match each character. The process includes three steps: (1) Extract all speakers from the story and create captions, (2) Create Facial personas for each character, and (3) Create audio personas for each character.
\subsubsection{Extract all speakers and their characteristics from story}
We first input the entire story into the LLM and extract all distinct characters with speaking lines, including the narrator. The LLM also makes captions about the physical appearance and personality traits of each character. Even if the story does not explicitly describe their appearance or intonation, we guide the LLM to infer these details through reasonable imagination based on indirect descriptions in the narrative. An example prompt is as follows :

\begin{tcolorbox}[colback=gray!10, colframe=black, boxrule=0.5pt, arc=4pt, width=\columnwidth]
\texttt{Analyze the following story and extract:}

\texttt{1. The narrator: provide a detailed description of the narrator's external features that could be used to create a portrait. If the narrator is a third-person narrator, analyze the main character in the story and treat them as the narrator.}

\texttt{2. A list of characters with speaking lines (dialogue) and for each character, provide their name or role in the story and a description of their external features that could be used to create a portrait in one sentence. If sufficient information is not provided, use reasonable imagination to infer their features based on their role and context.})
\end{tcolorbox}

\subsubsection{Create facial personas}

Using the captions generated in the previous step, we create AI-generated face images with a State-of-the-Art text-to-image model, the Stable Diffusion Model. Specifically, considering that the primary  dataset used to train the backbone TTS model is sourced from real human speech scenes, we employ a model specialized in generating photorealistic human images\footnote{The model is able to freely download from \url{https://huggingface.co/SG161222/Realistic_Vision_V2.0}}. Additionally, we filter out any samples that do not depict a human face.

\subsubsection{Create audio personas}
Using the face images and captions generated in the previous step, we create audio samples that match each character. Specifically, we use the Masked Generation feature of FleSpeech to synthesize short voice samples for each character using only face images and captions.

\begin{figure*}[ht]
    \centering
    \includegraphics[width=\textwidth]{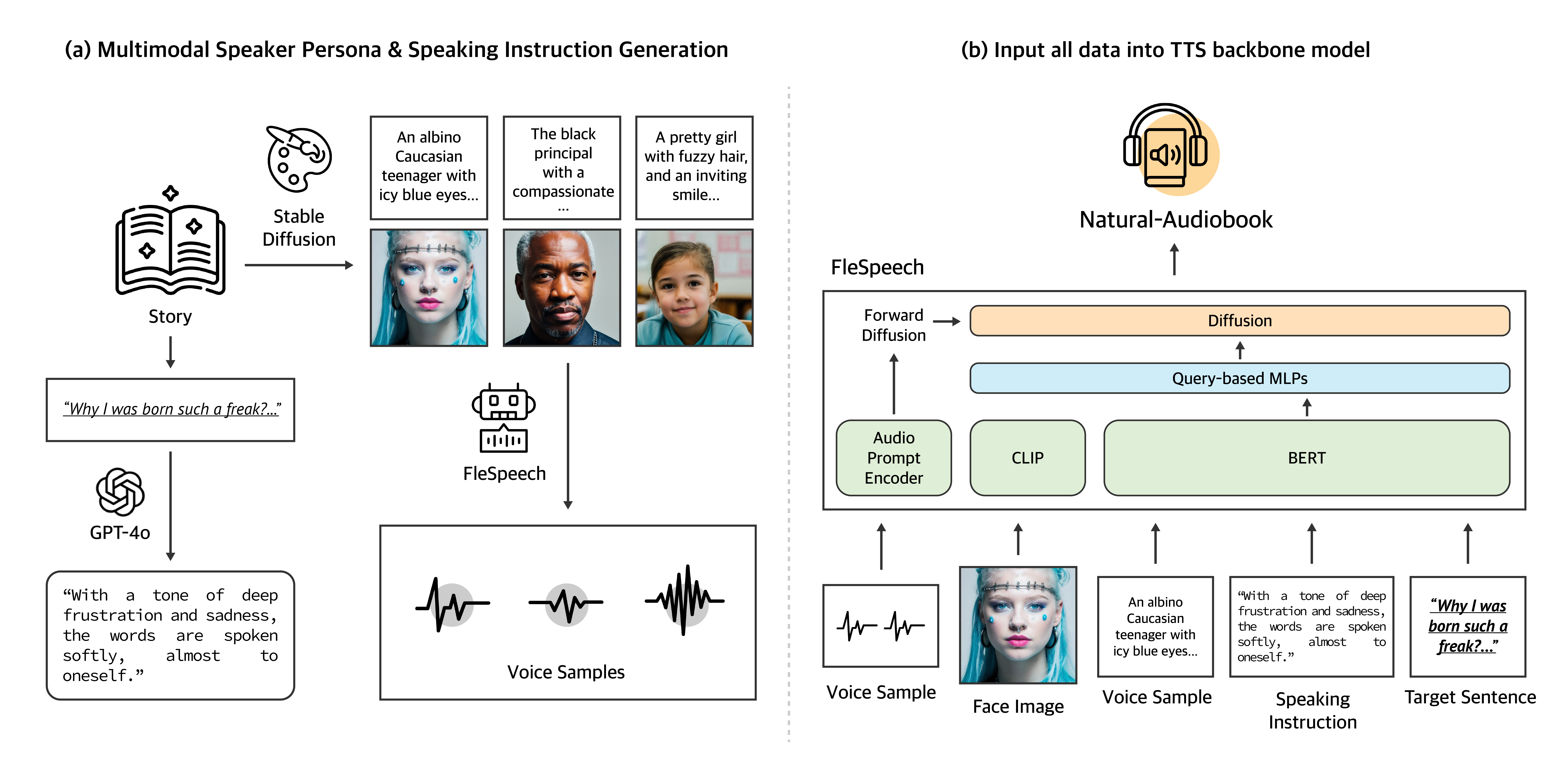} 
    \caption{\textbf{The MultiActor-Audiobook}. At the left side of figure (a), we perform Multimodal Speaker Persona Generation to create each speaker's AI-generated face images and voice samples, and LLM-based Script Instruction Generation to annotate every sentence's speaking instructions. At the right side of figure (b), we input all the multimodal input to our backbone TTS model, the FleSpeech, to generate speaker-aligned emotional audiobook.}
    \label{fig:wide_image}
\end{figure*}

\subsection{LLM-Based Script Instruction Generation}
In this process, we generate individual instructions for every sentence in the story. This process consists of two steps: (1) Identifying the speaker of each sentence, (2) Generating appropriate emotional instructions for each sentence.
\subsubsection{Identify the speaker of each sentence}
In this process, we input the entire story into the LLM to identify which character speaks each sentence. Since many dialogue lines are spoken by characters other than the narrator, accurately distinguishing the speaker of each line is crucial for the reader's experience. Once this process is complete, we match each sentence with the correct speaker's ID and persona information. An example prompt is as follows : 

\begin{tcolorbox}[colback=gray!10, colframe=black, boxrule=0.5pt, arc=4pt, width=\columnwidth]
\texttt{You are analyzing a story. Based on the full story and the current sentence, determine who is speaking the current sentence. 
    Use the following rules:
    1. Refer to the full context of the story to identify the speaker.
    2. Only assign the speaker from the provided list of characters below.
    3. Consider dialogue attribution markers (e.g., 'said Alice') and indirect clues in the story.
    4. If the sentence seems to be narration, return the main character's name.}
\end{tcolorbox}

\subsubsection{Create Text Description of Each Sentence}
In this process, we input the entire story, the target sentence, and the speaker information of the target sentence into the LLM to generate appropriate narration descriptions (e.g., \textit{Use a calm and reassuring tone.}). We instruct the model to avoid descriptions unrelated to speaking and to ensure that the emotional flow transitions naturally from the previous sentence to prevent abrupt emotional shifts. An example prompt is as follows:

\begin{tcolorbox}[colback=gray!10, colframe=black, boxrule=0.5pt, arc=4pt, width=\columnwidth]
\texttt{[Full Story] [Target Sentence] Based on the full context of the story and the current sentence, provide a single, concise instruction on how the sentence should be read aloud emotionally. Make the tone, pitch, pacing, and emotional delivery needed for a professional narration. If applicable, make a smooth transition from the previous emotion, and consider the emotions between the characters.}
\end{tcolorbox}

\subsection{Integration All Input Data}

Finally, we input all the data generated from the previous two processes—face images, face captions, audio samples, text descriptions, and target texts—into FleSpeech \cite{li2025flespeech}, the backbone multimodal TTS system, to complete the audiobook on a sentence-by-sentence basis. FleSpeech includes a unified multimodal prompt encoder, which is trained to map representations from various modalities using query-based MLPs and a diffusion process, allowing us to input text, audio, and visual data all at once. The face images, face captions, and audio samples remain fixed for each character within a single story sample generation, while only the target text and text description are updated for each sentence.

\begin{table*}[t]
    \centering
    \renewcommand{\arraystretch}{1.2}
    \begin{tabular}{lcccccc}
        \toprule
        & \textbf{Ours} &  w/o \texttt{MSP} &  w/o \texttt{LSI}  & ElevenLabs & FakeYou & F5-TTS \\
        \midrule
        Char-Con & 2.9$_{\pm0.12}$ / \textbf{3.9} & 2.6$_{\pm0.17}$ / 3.4 & 2.9$_{\pm0.10}$ / \underline{3.8} & \textbf{4.2}$_{\pm0.60}$ / 3.7 & \underline{3.2}$_{\pm0.31}$ / 3.2 & 2.6$_{\pm0.21}$ / 3.7 \\
        MOS-Q & 2.4$_{\pm0.29}$ / 3.3 & 2.2$_{\pm0.21}$ / 3.1 & 2.8$_{\pm0.20}$ / 3.4 & \textbf{4.6}$_{\pm0.27}$ / \textbf{4.0} & \underline{3.6}$_{\pm0.25}$ / \underline{3.7} & 3.4$_{\pm0.25}$ / 3.7 \\
        MOS-E & \underline{2.9}$_{\pm0.21}$ / \textbf{4.6} & 2.5$_{\pm0.19}$ / 4.0 & 2.6$_{\pm0.12}$ / 3.7 & \textbf{4.2}$_{\pm0.52}$ / 4.2 & 2.5$_{\pm0.27}$ / \underline{4.4} & 1.8$_{\pm0.31}$ / 3.9 \\
        MOS-S & \underline{2.6}$_{\pm0.24}$ / \textbf{3.4} & 2.4$_{\pm0.15}$ / \underline{3.3} & 2.5$_{\pm0.12}$ / 3.1 & \textbf{4.3}$_{\pm0.44}$ / 2.6 & 2.4$_{\pm0.23}$ / 3.2 & 2.0$_{\pm0.30}$ / 2.6 \\
        \bottomrule
    \end{tabular}
    \caption{\textbf{Human / MLLM average scores of} Char-Consistency(Char-Con), MOS-Quality(MOS-Q), MOS-Emotion(MOS-E), and MOS-Speaker(MOS-S) across our system and baselines. For each score, the value to the left of the
slash (/) represents the average MOS scores of human annotators, while the value to the right represents the
MOS scores of the gpt-4o-audio-preview. We bold the best values and underline the second best values. }
    \label{tab:comparison}
\end{table*}

\begin{table}[t]
    \centering
    \renewcommand{\arraystretch}{1.2}
    \begin{tabular}{lcc}
        \toprule
        & \textbf{Speaker Similarity} & \textbf{Turning Points} \\
        \midrule
        \textbf{Ours} & \textbf{51.334} & \textbf{146885.1} \\
        ElevenLabs & 40.473 & \underline{125309.6} \\
        FakeYou & 48.640 & 108087.5 \\
        F5-TTS & \underline{51.332} & 57737.7 \\
        \bottomrule
    \end{tabular}
    \caption{The speaker embedding similarities and number of turning points of our system and baselines. We bold the best values and underline the second best values. }
    \label{tab:pitch_std_turningpoints}
\end{table}

\section{Experiments}
\subsection{Experimental Setup}
\subsubsection{Story Dataset}

We use the story dataset ReedsyPrompts\cite{park2024longstory} in this paper. The ReedsyPrompts contains appropriate length of stories, allowing for diverse and distinctive speakers to appear within a single story. We use only partial samples of the total training samples since the heavy generation costs. In 12 samples, each story has an average of 4.3 speakers and consists of an average of 175 sentences. The average audio length is 749.56 seconds.


\subsubsection{MultiActor-Audiobook Implementation Details} 

For generating face images, we use the State-of-the-Art photorealistic Stable Diffusion model, the \textit{SG161222/Realistic\_Vision\_V2.0}. We also utilize gpt-4o-2024-08-06 for the LLM capabilities, and FleSpeech served as the backbone TTS model, with the same parameter values as reported in the original paper. None of the models  perform any additional training, and for the inference we use NVIDIA RTX A5000 GPU settings.

\subsubsection{Baselines} 

The baselines are ElevenLabs, FakeyouTTS, F5-TTS,  w/o  Persona, and  w/o  Instruction. ElevenLabs and FakeyouTTS are well-known commercial TTS systems that do not support the input of speaking descriptions or multimodal persona information for speakers. Instead, they offer predefined character intonations that users can manually select. Therefore, we choose the most suitable character intonation for each story. We generate each audiobook using the same stories as our system. Note that ElevenLabs supports audiobook generation in a life-like and emotionally rich mode.

F5-TTS allows specifying the speaking voice style of each character but does not utilize visual persona information or speaking descriptions. We use each character's audio persona to determine the voice style for each sentence.

 w/o \texttt{MSP} is a version that uses the same backbone TTS model of ours but masks the Multimodal Speaker Personas which include face images, face captions, and audio samples, relying solely on the target text and text descriptions.

 w/o \texttt{LSI} refers to a version that utilizes all input data except for the LLM-based Script Instruction.

\subsection{Main Experiments Results}


We conduct a Mean Opinion Score (MOS) listening test with five human evaluators and an MLLM evaluator. Character-Voice Consistency measures how well the audiobook’s voices align with the characters' personality in the story, including factors such as gender, pitch, speed, volume, and intonation. MOS-Q evaluates the overall audio quality, including aspects like clarity, high-frequency, and naturalness. MOS-E assesses how appropriately emotions are conveyed in each sentence, while MOS-S measures how accurately the speaker is identified for each sentence. For MLLM evaluations, we used the same question provided to human evaluators and employed the audio-to-text QA feature of gpt-4o-audio-preview.

The quantitative analysis involves the analysis of speaker embedding similarity and the number of pitch turning points. To evaluate the voice consistency of the audiobooks, we measure the speaker embedding every 10 seconds in a sample using an open-source speech embedding model\footnote{\url{https://github.com/douglas125/SpeechIdentity?tab=readme-ov-file}}, and calculate the average similarity. To assess emotional expressiveness, we count the number of pitch turning points. A higher speaker embedding similarity indicates better maintenance of voice consistency within the sample, while a greater number of pitch turning points suggests stronger emotional fluency.

\subsubsection{Character-Voice Consistency}
As shown in Table \ref{tab:comparison}, ElevenLabs achieves the highest Character-Consistency score in human evaluation, followed by FakeYou. In the MLLM evaluation, our system scores the highest, while  w/o \texttt{LSI} achieves the second-highest score. Although ElevenLabs' audiobook generation does not support automatic selection of an optimistic narrator voice, its substantial additional training cost may contribute to its superior performance compared to other baselines.

For FakeYou, we manually selected the best-matched famous actors' voices, such as Morgan Freeman, for each story, which likely led to strong human evaluator preferences in the Character-Consistency metric. However, the MLLM scores indicate that our system also performs well compared to other baselines, thanks to Multimodal Speaker Persona Generation.

\subsubsection{MOS-Q, MOS-E, MOS-S} 
As shown in Table \ref{tab:comparison}, ElevenLabs achieves the highest MOS-Q scores in both human and MLLM evaluations, followed by FakeYou. Compared to the zero-shot nature of our system, there commercial products have a significant advantage in this metric due to their substantial additional training costs.

For MOS-E and MOS-S scores, our system ranks second-highest in human evaluation and achieves the highest scores in MLLM evaluation. Its superior performance compared to  w/o \texttt{MSP} and  w/o \texttt{LSI} demonstrates that both of our novel processes—the Multimodal Speaker Persona Generation and LLM-based Script Instruction Generation—are crucial for conveying appropriate emotional expressions.

Since  w/o \texttt{LSI} outperforms  w/o \texttt{MSP} across all metrics, particularly in Char-Con and MOS-Q scores, we conclude that multimodal inputs play a central role in audiobook quality, primarily by enhancing the consistency of speaker identities.

\subsubsection{Quantitative Analysis}
As shown in Table \ref{tab:pitch_std_turningpoints}, our system achieves the highest Speaker Similarity and number of pitch turning points. From this, we can conclude that our system maintains voice consistency across each sample very well while also producing various levels of emotional expressions, such as joy, love, and laughter, which result in frequent pitch variations.

\section{Conclusion}
We introduce MultiActor-Audiobook, a zero-shot system for generating emotionally expressive and speaker-appropriate audiobook without extra training or manual annotations. Our approach leverages two key processes: \textbf{Multimodal Speaker Persona Generation}, and \textbf{LLM-Based Script Instruction Generation}. Experimental results show that our system achieves mostly best or second highest scores in human and MLLM evaluations, and the best results in quantitative analysis, while eliminating the need for costly data collection and manual labeling.

\section{Limitation}
MultiActor-Audiobook exhibits lower quality than some commercial systems, mainly due to the backbone TTS model, FleSpeech, which is trained on a smaller, less specialized dataset. Large-scale audiobook datasets by professional voice actors could improve performance. Additionally, since FleSpeech was trained on specific face samples (e.g., TED lecturers), it struggles with unseen AI-generated faces. Future multimodal TTS systems with more diverse face samples could enhance MultiActor-Audiobook’s performance.

\bibliographystyle{IEEEtran}
\bibliography{mybib}

\end{document}